\documentstyle[preprint,aps,eqsecnum]{revtex}

\def\red{
}
\def\black{
}

\def\mycomm#1{\hfill\break\strut\kern-3em{\red\tt ====> #1\black}\hfill\break}
\def\mycommNL#1{\strut\kern-3em{\red\tt ====> #1\black}\hfill\break}
\def\ds{\displaystyle}
\def\deepstrut{\vrule height 1.5ex depth 3.5ex width 0pt}

\def\verydeepstrut{\vrule height 1.5ex depth 4.5ex width 0pt}
\def\tallstrut{\vrule height 4.5ex depth 1.5ex width 0pt}

\def\understrut{\vrule height 2.5ex depth 1.0ex width 0pt}
\def\substrut{\vrule height 1.0ex depth 0ex width 0pt}

\def\eqref#1{(\ref{#1})}

\def\MUU{ {\cal V } }
\def\MUD{ {\cal P } }

\makeatletter
\def\hlinewd#1{\noalign{\ifnum0=`}\fi
\hrule \@height #1 \futurelet \reserved@a\@xhline}
\def\hwhiteline{\noalign
{\ifnum0=`}\fi\hrule
\@height 0pt\vskip 1.0ex\futurelet \reserved@a\@xhline}
\makeatother
\def\gray{\special{ps: 0.40 setgray}}
\def\black{\special{ps: 0.0 setgray}}

\newcommand{\mydraft}{
\newcount\timecount
\newcount\hours \newcount\minutes  \newcount\temp \newcount\pmhours

\hours = \time
\divide\hours by 60
\temp = \hours
\multiply\temp by 60
\minutes = \time
\advance\minutes by -\temp
\def\hour{\the\hours}
\def\minute{\ifnum\minutes<10 0\the\minutes
    \else\the\minutes\fi}
\def\clock{
\ifnum\hours=0 12:\minute\ AM
\else\ifnum\hours<12 \hour:\minute\ AM
\else\ifnum\hours=12 12:\minute\ PM
    \else\ifnum\hours>12
     \pmhours=\hours
     \advance\pmhours by -12
     \the\pmhours:\minute\ PM
     \fi
    \fi
\fi
\fi
}
\def\fullclock{\hour:\minute}
\begin{centering}
\gray
\font\Hugett  =cmtt12 scaled\magstep4
\hbox{\Hugett Draft:\today,\clock}
\black
\end{centering}
\vskip -1.7cm
$\phantom{a}$
} 

\def\beq#1{\begin{equation} \label{#1}}
\def\eeq{\end{equation}}

\newskip\humongous \humongous=0pt plus 1000pt minus 1000pt

\newif\ifdtup

\begin{document}
{\tighten
\preprint
{\vbox{
\hbox{TAUP 2829/06}
\hbox{WIS/07/06-JULY-DPP}
\hbox{ANL-HEP-PR-06-56}
}}

\title{New Quark Relations for Hadron Masses and Magnetic Moments \\
A Challenge for Explanation from QCD}
\author{Marek Karliner\,$^{a}$\thanks{e-mail: \tt marek@proton.tau.ac.il}
\\
and
\\
Harry J. Lipkin\,$^{a,b}$\thanks{e-mail: \tt
ftlipkin@weizmann.ac.il} }
\address{ \vbox{\vskip 0.truecm}
$^a\;$School of Physics and Astronomy \\
Raymond and Beverly Sackler Faculty of Exact Sciences \\
Tel Aviv University, Tel Aviv, Israel\\
\vbox{\vskip 0.0truecm}
$^b\;$Department of Particle Physics \\
Weizmann Institute of Science, Rehovot 76100, Israel \\
and\\
High Energy Physics Division, Argonne National Laboratory \\
Argonne, IL 60439-4815, USA\\
}
\maketitle
\begin{abstract}%

Prompted by the recent surprising results in  QCD spectroscopy, we
extend the treatment of the constituent quark model showing that
mass differences and ratios have the same values when  obtained
from mesons and baryons. We obtain several new successful
relations involving hadrons containing two and three strange
quarks and hadrons containing heavy quarks
and give a new prediction regarding spin splitting between
doubly charmed baryons.
We provide numerical evidence for an effective supersymmetry between 
mesons and baryons related by replacing a light antiquark by a light diquark.
 We also obtain new
relations between quark magnetic moments and hadron masses. Limits
of validity of this approach  and disagreements with experiment in
properties of the $\Sigma$ and $\Xi$ baryons are discussed as
possible clues to a derivation from QCD.
\end{abstract}%
} 

\section{Introduction }

\subsection{What is a constituent quark?}

Nature tells us in experimental data that mesons and baryons are
made of the same building blocks, sometimes called
``constituent quarks".
Mesons are two blocks and nothing else,
baryons are three blocks and nothing else, and no present theory
tells us what they are.

The challenge for QCD is to explain the structure of these blocks in
quarks, antiquarks and gluons and why they are the same in mesons and
baryons.

Early evidence that mesons and baryons are made of the same building
blocks appeared in the remarkable successes of the constituent quark model.
Static properties, low lying excitations and total scattering cross
sections of both mesons and baryons are described as simple composites of
asymptotically free quasiparticles with given effective masses
\cite{SakhZel,ICHJLmass,HJLMASS,sigtot,nuhyp}.

The last few years have brought a rich crop of surprises in QCD
spectroscopy \cite{Rosner:2006jz}. These include too many experimental
results relating mesons and baryons to be an accident. Their explanation
remains a challenge for
QCD\cite{Godfrey:1985xj,Capstick:1986bm,Manohar:1983md}. Some of the new
states seen have not been predicted at all; others are exceedingly narrow
with properties very different from most theoretical expectations. This
has prompted us to re-examine several aspects of the constituent quark
model, to extend the experimental basis for simple meson-baryon relations
and to search for clues to the eventual description by defining the domain
where the simple model succeeds and where it fails.

Extending these mesons-baryon relations to include heavy quarks shows that
the simple constituent quark relations hold in some cases and break down
in others. They hold between mesons that are bound states of a quark of
any flavor and a color antitriplet light antiquark fermion, and baryons
that are bound states of a quark of the same flavor and a color
antitriplet light diquark boson, for all quark flavors. They seem to break
down for states containing more than one heavy or strange quark.

Thus the QCD interaction between a color triplet heavy quark and a color
antitriplet light quark system appears not to be sensitive to the
structure of the light quark system; i.e. whether the color antitriplet is
an antiquark or a color antitriplet $ud$ pair. This suggests some kind of
effective supersymmetry between hadrons related by replacing
a light fermion by a
light boson.
The question arises whether these relations are obtainable
from any of the known approaches to QCD or come from an effective light
quark supersymmetry which is yet to be derived from QCD.
So far the experimental evidence is impressive, and none of the
various approaches to QCD seem to incorporate this symmetry.
This is a very exciting challenge for theory.

The relation between the {\em constituent quarks} and the
fundamental fields appearing in the QCD Lagrangian, the {\em
current quarks}, remains to be understood. Perhaps there is no such
relation and the success of the  constituent quark model in relations
between mesons and baryons is only a key to a hidden diquark-antiquark
symmetry or effective supersymmetry.

Until now, lattice QCD is the only theoretical approach which
starts from the fundamental fields of QCD and computes the spectrum.
Despite this, many phenomenological relations between observables are hard
to understand within the framework of lattice QCD, while they appear
natural in the constituent quark model. This is why the elucidation of the
relation between the effective and fundamental degrees of freedom is so
important.

The obvious approach of treating a constituent quark as a current quark,
valence quark or ``bare quark" surrounded by a cloud or ``sea" of gluons,
$q\bar q$ pairs or pions has been tried many times and failed. A major
difficulty is explaining how the same cloud works for the constituent
quarks in both mesons and baryons.
There are also sea quark effects
which are known to be important for magnetic moments \cite{Leinweber:2002qb}.
The
constituent quarks somehow automatically incorporate such effects.
What is missing and what we are unable
to do at this stage is a theoretical
derivation of such effects {\em from first principles} and their incorporation
into the quark model.

Gell-Mann has suggested that
constituent quarks are related to current quarks by a unitary
transformation. However no such unitary transformation has been found. It
may well be as complicated as the transformation between the electrons in
QED and the quasiparticles needed to explain the fractional quantum Hall
effect. Or it may not exist at all and merely be manifestation of a hidden
effective supersymmetry.

We search for further
illumination on this question by pursuing the successes and failures of
the simple constituent quark model in unambiguous predictions of experimental
data which can be clearly shown to be either right or wrong without adjusting
free parameters.

\subsection{Some Simple Successes}

The successes of the constituent quark model in explaining regularities in
experimental data that are not explained by other approaches are already
too extensive to be dismissed as accidental. For example,
calculations from experimental baryon masses and from meson masses give
the same values $\pm 3\%$ for the effective quark mass difference
$m_s-m_u$ between the strange and up quarks and their mass ratio
$m_s/m_u$. QCD calculations have not yet succeeded to explain these
striking experimental facts. The search for some QCD model for the
structure of the constituent quark or a unitary transformation or
effective supersymmetry is therefore of interest.

We search for clues to this structure or transformation by extending the
domain where the simple model works as far as possible, while noting also
the limits of its validity. One remarkable success of this model is its
prediction\cite{Protvino} of the absolute value of the isoscalar nucleon
magnetic moment\cite{PDG} with no free parameters.

\beq{isomag1}
\mu_p+\mu_n= 2M_{\scriptstyle p}\cdot {{Q_I}\over{M_I}} ={2M_N\over
M_N+M_\Delta}=0.865 \,{\rm n.m.} \qquad(\hbox{EXP} = 0.88 \,{\rm n.m.})
\end{equation}
where $ \ds Q_I= {1\over 2}\cdot \left( {2\over3} -
{1\over 3} \right) = {1\over 6} $ \ and \ $ \ds M_I= {1\over 6}\cdot
\left( M_N + M_\Delta \right)$ denote the charge and mass, respectively,
of an effective ``isoscalar nonstrange quark".

This simple derivation of the isoscalar nucleon moment is
remarkable for
giving an absolute prediction, not
merely a ratio.
It
sets a mass scale
in remarkable agreement with experiment by
simply stating that the isoscalar nucleon magnetic moment is the Dirac
moment of an isoscalar quark with a charge of (1/6) and a mass
(1/3) of the mean mass of the nucleon and the $\Delta$; i.e. the mass
of a three quark system with the hyperfine energy removed. This value
for an ``effective" quark mass originally proposed by Sakharov and
Zeldovich\cite{SakhZel} has led to many successful relations between
hadron masses\cite{SakhZel,ICHJLmass,HJLMASS}.

This one prediction assumes no specific spin couplings of the quarks; e.g.
$SU(6)$, as in the ratio relations\cite{DGG}. The total spin contribution
to the magnetic moment of a system of three identical quarks coupled to
total spin 1/2 is rigorously equal to the magnetic moment of a single
quark. Why this works so well and how this scale arises from a real theory
is a challenge for QCD and is not easily dismissed as an accident.

Constituent quark predictions for the proton, neutron and $\Lambda$
magnetic moments follow of a baryon model of three constituent quarks and
nothing else with Dirac moments having effective masses determined
uniquely from hadron masses.
The success of this description implies a complicated structure for the
constituent quarks. The physical proton is known to consist of three
valance quarks, a sea of quark-antiquark pairs and gluons. That all these
constituents can be described so well by three constituent quarks and
nothing else and give such remarkable agreement with experiment is a
mystery so far unexplained by QCD. Continuing this approach leads to
remarkable agreement with experiment following from the assumption that
all ground state baryons are described by three constituent quarks and
nothing else and that all ground state mesons are described by quark
antiquark pairs and nothing else and that the constituent quarks in mesons
and baryons are the same.

A completely different experimental confirmation of this picture is seen in the
relations between meson-nucleon and baryon-nucleon total cross
sections\cite{sigtot,nuhyp}. One example is the successful predictions for
baryon-nucleon total cross sections from meson-nucleon cross sections
at $P_{lab}$ = 100 GeV/c,
\beq{sigtotp}
38.5 \pm 0.04 \,{\rm mb} = \sigma_{tot}(pp) =
3 \sigma_{tot}(\pi^+p) -{3\over 2}\sigma_{tot}(K^-p) =39.3 \pm 0.2 \,{\rm mb}
\end{equation}
\beq{sigtotsig}
33.1 \pm 0.31 \,{\rm mb} = \sigma_{tot}(\Sigma p) =
{3\over 2}\{\sigma_{tot}(K^+p) + \sigma_{tot}(\pi^-p) -\sigma_{tot}(K^-p)\}
=33.6 \pm 0.16 \,{\rm mb}
\end{equation}
\beq{sigtotxi}
29.2 \pm 0.29 \,{\rm mb} = \sigma_{tot}(\Xi p) =
{3\over 2}\sigma_{tot}(K^+p) =28.4 \pm 0.1 \,{\rm mb}
\end{equation}
But we still do not know what the constituent
quark is.

\subsection{The search for further clues}

We continue this search for clues to the nature of the constituent quark
by presenting here new relations between meson and baryon masses that are
in surprising agreement with experiment. On the other hand we sharpen the
disagreement between the experimental values of hyperon magnetic moments
and the predictions of this simple picture.

We find relations between masses of mesons and baryons
containing two or more strange quarks which confirm this picture. This
is made
possible by assuming that the spin dependence of the $qq$ and $\bar qq$
interactions is that of a hyperfine interaction with the ratio of
1/(-3)
between triplet and singlet states, and that the interaction is
proportional
to the product of the quark color magnetic moments which are
proportional to
the quark electromagnetic magnetic moments determined from the measured
nucleon and $\Lambda$ moments.

We also examine new relations involving masses of hadrons containing
heavy
quarks. The same approach used for light hadrons leads to successful
mass
relations between hadrons containing one heavy quark and light u and d
diquarks or antiquarks.

The approach breaks down for states containing additional
heavy or strange quarks or antiquarks. This suggests that the
distance between two heavy quarks can be sufficiently small to be
in the range of the strong short-range coulomb-like force. In
states containing only one heavy quark the low mass of the rest of
the system produces a low reduced mass and therefore a high
kinetic energy for localization in the domain of the coulomb-like
interaction. We also present cases where the simple model fails,
since the contrast between extraordinary success and failure can
provide clues to the more fundamental derivation from QCD.

\section{New mass relations between masses of hadrons containing two strange
quarks}

The first suggestion that hadron spin splittings  arise from a $qq$ and $\bar
q q$ hyperfine interaction that is the same in mesons and baryons was due to
Andrei Sakharov, a pioneer in quark-hadron physics. He asked in 1966 ``Why are
the $\Lambda$ and $\Sigma$ masses different? They are made of the same quarks".
Sakharov and  Zeldovich\cite{SakhZel}  assumed a $universal$ quark model for both
mesons and baryons with
a flavor dependent linear mass term and hyperfine interaction. The success in fitting
experiment of this
Sakharov-Zeldovich
universality which relates meson and baryon masses with the same quark mass parameters
remains a challenge to QCD. So far
all QCD
treatments tend to treat meson and baryon structures very differently.

The updated\cite{DGG} version of the
Sakharov-Zeldovich mass formula\cite{SakhZel} is

\beq {sakzel}
M = \sum_i m_i + \sum_{i>j}  {{\vec{\sigma}_i\cdot\vec{\sigma}_j}\over{m_i\cdot
m_j}}\cdot v^{hyp}_{ij}
\end{equation}
where $m_i$ is the
effective mass\cite{Postcard} of quark $i$, $\vec{\sigma}_i$ is a quark  spin operator and
$v^{hyp}_{ij}$ is a hyperfine interaction with different strengths but the same
flavor dependence.

Eq.~\eqref{sakzel} yields to the following expressions for baryon masses
\begin{equation}
\matrix{
M_N &=&
\ds 3 m_u - 3 V_{hyp}(uu)
\verydeepstrut
\cr
M_{\Delta}&=&
\ds 3 m_u + 3 V_{hyp}(uu)
\verydeepstrut
\cr
M_{\Lambda}&=&
\ds 2 m_u + m_s -3 V_{hyp}(uu)
\verydeepstrut
\cr
M_\Sigma &=&
\ds 2 m_u + m_s + V_{hyp}(uu) - 4 V_{hyp}(us)
\verydeepstrut
\cr
M_{\Sigma^{\scriptstyle *}} &=&
\ds 2 m_u + m_s + V_{hyp}(uu) + 2 V_{hyp}(us)
\verydeepstrut
\cr
M_\Xi &=&
\ds 2 m_s + m_u + V_{hyp}(ss) - 4 V_{hyp}(us)
\verydeepstrut
\cr
M_{\Xi^{\scriptstyle *}} &=&
\ds 2 m_s + m_u + V_{hyp}(ss) + 2 V_{hyp}(us)
\verydeepstrut
\cr
M_{\Omega^{\scriptstyle -}}&=&
\ds 3 m_s + 3 V_{hyp}(ss)
\verydeepstrut
}
\label{baryon_masses}
\end{equation}

\vskip-2cm

An immediate and well known consequence of the universality of
eq.~(\ref{sakzel}) is
a relation showing that
the mass difference $m_s-m_u$ has the same value\cite{PDG}
when calculated from baryon masses and meson masses\cite{SakhZel}.
\def\mykern{\kern-0.3em}
\begin{equation}
\kern-0.5em
\matrix{
\langle m_s{-}m_u \rangle_{Bar}
\mykern
&=&
\mykern
M_{sud}{-}M_{uud}
\mykern
&=&
\mykern
M_\Lambda{-}M_N
\mykern
&=&
\mykern
177\,{\rm MeV} \deepstrut
\cr
\langle m_s{-}m_u \rangle_{Mes}
\mykern
&=&
\mykern
\ds
{
3  (M_{\MUU_{s\bar d}} {-} M_{\MUU_{u\bar d}})
{+}(M_{\MUD_{s\bar d}} {-} M_{\MUD_{u\bar d}})
\understrut
\over 4}
\mykern
&=&
\mykern
\ds { 3(M_{K^{\scriptstyle *}}{-}M_\rho){+}M_K{-}M_\pi \over 4 }
\mykern
&=&
\mykern
178\,{\rm MeV}
\verydeepstrut
}
\label{SZeq}
\end{equation}
where the {\em ``Bar"} and {\em ``Mes"} subscripts denote values obtained
from baryons and mesons, and
$\MUU$ and $\MUD$ denote vector and pseudoscalar mesons, respectively.

The original Sakharov-Zeldovich mass formula has no mass dependence
in the hyperfine interaction.
The mass dependence was introduced later\cite{DGG}
and used to derive the $\Lambda$ moment. A weaker version
\cite{Protvino} which did not use a mass dependence used a
hyperfine interaction proportional to the
product of quark color magnetic moments which are in turn proportional to
quark electromagnetic moments.
     The relation between these three versions of eq.~(\ref{sakzel}) in
fitting
hadron masses and magnetic moments has been studied in
detail\cite{NewPenta}. The question of whether the quark masses appearing
in the two terms in eq.~(\ref{sakzel}), the additive mass term and the
hyperfine
interaction, should have the same values in both terms has been analyzed
in fits to all previously available data. We go beyond this
treatment\cite{NewPenta} in looking for new relations not previously
considered while keeping track of which of the three versions of
eq.~(\ref{sakzel}) is used in each case.

We first obtain interesting new relations from  the assumption
that the  $qq$ and $\bar q q$  hyperfine interactions are proportional to the
product of  the color magnetic moments of the quarks without assuming a
specific mass dependence and then relate this to the quark electromagnetic
moments.

Let $V_{hyp}(uu)$, $V_{hyp}(us)$ and
$V_{hyp}(ss)$  denote the color magnetic energies respectively in the $uu$,
$us$ and $ss$ systems in states of spin 1 and assume that the spin dependence
of the interaction is that of a hyperfine  interaction with the ratio of 1/(-3)
between triplet and singlet states.
The ratio   $[V_{hyp}(uu) -V_{hyp}(ss)]/V_{hyp}(uu)$ can then be obtained in
two different ways from the baryon spectrum,
\beq{dechyp}
{{M_\Delta - M_{\Omega^-} +  3[M_\Lambda-M_N]}\over{M_\Delta - M_N}}
 =
 {{3\cdot [V_{hyp}(uu) -V_{hyp}(ss)]}\over{6 V_{hyp}(uu)}}= 0.31
\end{equation}

\beq{declinear2}
{{3 [M_{\Sigma^{\scriptstyle *}} - M_{\Xi^{\scriptstyle *}} +
M_\Lambda-M_N]}\over{M_\Delta - M_N}}
 =
 {{3\cdot [V_{hyp}(uu) -V_{hyp}(ss)]}\over{6 V_{hyp}(uu)}}= 0.28
\end{equation}
where the l.h.s. of eqs.~\eqref{dechyp} and \eqref{declinear2} are
 constructed so that the contributions of
the quark mark differences cancel.

Combining (\ref{dechyp}) and (\ref{declinear2}) gives
a precise relation between decuplet hyperfine splittings, good to 2\%,
\beq{decuplet}
\matrix{
M_{\Xi^{\scriptstyle *}}-M_{\Sigma^{\scriptstyle *}} &=&
\ds (m_s - m_u) - V_{hyp}(uu) + V_{hyp}(ss)
&=& 149 \,{\rm MeV}
\verydeepstrut
\cr
\displaystyle {{M_{\Sigma^{\scriptstyle *}}-  M_{\Delta}}\over 2} +
{{M_{\Omega^{\scriptstyle -}} - M_{\Xi^{\scriptstyle *}}}\over 2} &=&
\ds (m_s - m_u) -  V_{hyp}(uu) +  V_{hyp}(ss)
&=& 146 \,{\rm MeV}
\verydeepstrut
}
\end{equation}
The importance of the flavor
dependence of the hyperfine interaction is seen in the contrast between
this agreement and an analogous test of the ``equal
spacing rule" for $SU(3)$ breaking in the decuplet which neglects this flavor
dependence and is good only to 10\%,
\deepstrut
\beq{declinear}
\matrix{
M_{\Sigma^{\scriptstyle *}}-  M_{\Delta} &=&
\ds (m_s - m_u) - 2 V_{hyp}(uu) + 2 V_{hyp}(us)
&=& 153 \,{\rm MeV}
\verydeepstrut
\cr
M_{\Omega^{\scriptstyle -}} - M_{\Xi^{\scriptstyle *}}-&=&
\ds  (m_s - m_u) +  2V_{hyp}(ss) - 2 V_{hyp}(us)
&=& 139 \,{\rm MeV}
\verydeepstrut
}
\end{equation}

If hyperfine splittings in mesons are also proportional
to the analogous product of color magnetic moments, then

\beq{ssmes}
{{2[M_\rho - M_\phi] +  4[M_\Lambda-M_N]}\over{M_\rho - M_\pi}}=
 {{2\cdot [V_{hyp}(u\bar u) -V_{hyp}(s\bar s)]}\over{4 V_{hyp}(u\bar u)}}= 0.35
\end{equation}
where $V_{hyp}(u \bar u)$ and
$V_{hyp}(s \bar s)$ denote the color magnetic energies respectively $u \bar u$
and $s \bar s$ systems in states of spin 1 and we assume that the flavor
$SU(3)$
breaking factor has the same value for $q \bar q$ and $qq$ hyperfine
interactions

\beq{qqbar}
{{V_{hyp}(s\bar s)}\over{V_{hyp}(u\bar u)}}=
{{V_{hyp}(ss)}\over{V_{hyp}(uu)}}
\end{equation}
Then
\beq{sshyp3}
{{M_\Delta - M_{\Omega^-} +  3[M_\Lambda-M_N]}\over{M_\Delta - M_N}}
 = 0.31 \quad \approx \quad
{{2[M_\rho - M_\phi] +  4[M_\Lambda-M_N]}\over{M_\rho - M_\pi}}=
0.35
\end{equation}

If we assume that the color magnetic moments and the electromagnetic moments of
the quarks are proportional to one another\cite{DGG} we obtain

\beq{decmag} {{M_\Delta - M_{\Omega^-} +  3[M_\Lambda-M_N]}\over{M_\Delta -
M_N}} = {{3\cdot [V_{hyp}(uu) -V_{hyp}(ss)]}\over{6 V_{hyp}(uu)}}= 0.31 =
{{1}\over{2}}\cdot \left[1 -
\left({{\mu_s}\over{\mu_d}}\right)^2\right]
\end{equation}
\beq{decmag2}
{{3 [M_{\Sigma^{\scriptstyle *}} - M_{\Xi^{\scriptstyle *}} +
M_\Lambda-M_N]}\over{M_\Delta - M_N}}
 =
 {{3\cdot [V_{hyp}(uu) -V_{hyp}(ss)]}\over{6 V_{hyp}(uu)}}= 0.28 =
{{1}\over{2}}\cdot \left[1 -
\left({{\mu_s}\over{\mu_d}}\right)^2\right]
\end{equation}
This gives two values for the quark magnetic moment ratio obtained from baryon
masses

\beq{magrat}
\left({{\mu_d}\over{\mu_s}}\right)_{\tallstrut\stackrel{\Delta\,\Omega}{\substrut mass}}
\kern-0.5em
= 1.61; \qquad\,\,\,
\left({{\mu_d}\over{\mu_s}}\right)_{\tallstrut\stackrel{\Sigma^*\,\Xi^*}{\substrut
mass}} = 1.52; \qquad\,\,\,
\hbox{average:}\quad
\left({{\mu_d}\over{\mu_s}}
\right)_{\tallstrut\stackrel{\hbox{\footnotesize \it ss}}{\substrut
mass}} = 1.57
\end{equation}
to be compared to the ratio obtained\cite{DGG} from measured magnetic moments
\deepstrut
\beq{magratEXP}
\left({{\mu_d}\over{\mu_s}}\right)_{mag}
= {-}{{\mu_p}\over{3\mu_{\Lambda}}} =1.54
\end{equation}

\section{ Relations between masses of baryons and mesons containing
light quarks}

The relation
(\ref{sshyp3})
has been obtained only by relating the flavor dependences of the hyperfine
interactions in mesons and baryons without relating them to quark masses.
Including the explicit dependence on the quark masses\cite{DGG} now gives

\beq{hypmass} {{M_\Delta - M_{\Omega^-} +  3[M_\Lambda-M_N]}\over{M_\Delta -
M_N}} = {{3\cdot [V_{hyp}(uu) -V_{hyp}(ss)]}\over{6 V_{hyp}(uu)}}= 0.31 =
{{1}\over{2}}\cdot \left[1 -
\left({{m_u}\over{m_s}}\right)_{Bar\:ss}^2\right]
\end{equation}

\beq{mesmass}
{{2[M_\rho - M_\phi] +  4[M_\Lambda-M_N]}\over{M_\rho - M_\pi}}=
{{2\cdot [V_{hyp}(u \bar u) - V_{hyp}(s \bar s)}]
\over{4 V_{hyp}(u \bar u)}}=0.35
={{1}\over{2}}\cdot \left[1 -
\left({{m_u}\over{m_s}}\right)_{Mes\:ss}^2\right]
\end{equation}

\tallstrut Solving these equations for the ratio ${{m_s}/{m_u}}$ 
gives two values obtained respectively for mesons and baryons from the ratio of
the  doubly-strange hyperfine interaction to the corresponding
nonstrange interaction,

\beq{masrat}
\left({{m_s}\over{m_u}}\right)_{Bar\:ss} = {{V_{hyp}(ss)}
\over{V_{hyp}(uu)}} = 1.61; ~ ~ ~
\left({{m_s}\over{m_u}}\right)_{Mes\:ss} =  {{V_{hyp}(s \bar s)}
\over{V_{hyp}(u \bar u)}}  = 1.80 
\end{equation}

These can be compared with the value for
${{m_s}/{m_u}}$ obtained  from the ratio of the  singly strange hyperfine
interaction $V_{hyp}(su)$ to the corresponding nonstrange $V_{hyp}(uu)$ as
given by eq.~(\ref{sakzel}),

\begin{equation} \left({{m_s}\over{m_u}}\right)_{Bar} = {{M_\Delta -
M_N}\over{M_{\Sigma^*} - M_\Sigma}} = 1.53 \quad\approx\quad
\left({{m_s}\over{m_u}}\right)_{Mes} = {{M_\rho - M_\pi}\over{M_{K^*}-M_K}}=
1.60 \verydeepstrut \label{SZmag} \end{equation}

The 5\% difference between quark mass ratios obtained from meson and baryon
masses is sufficiently small to provide a challenge to models of QCD. The model-dependent
explanations for this difference in simple potential models\cite{ICHJLmass} are
beyond the scope of the present treatment. They give a flavor-independent
difference of about 50 MeV between the effective quark masses of mesons and
baryons. This provides the needed correction to relations between effective
quark mass ratios while not affecting mass differences and explains why these
corrections are smaller for  heavier quarks.

The increase in the value of $m_s/m_u$ with decreasing hadron radius is
seen to continue monotonically also when the doubly strange mesons and
baryons are included.

\begin{equation}
\left({{m_s}\over{m_u}}\right)_{Bar} <
\left({{m_s}\over{m_u}}\right)_{Mes} <
\left({{m_s}\over{m_u}}\right)_{Bar\:ss}<
\left({{m_s}\over{m_u}}\right)_{Mes\:ss}; ~ ~ ~
\langle r^2 \rangle _\Delta>
\langle r^2 \rangle _ \rho>
\langle r^2 \rangle _\Omega >
\langle r^2 \rangle _\phi
\end{equation}
where $\langle r^2 \rangle$ denotes the mean square distance between two
constituents in the hadron.

\section {New relations between meson and baryon masses from hadrons
containing heavy quarks.}

We now continue to generalize eq.~(\ref{sakzel}) to other flavors
where excellent results have already been obtained\cite{PBIGSKY}.
Consider
hadrons containing a quark \strut \ $q_i$ \ or \ $q_j$ and
a ``spectator" color antitriplet $\bar x$.
Thus, given the masses \ of \ two vectors
\break
\ $|\MUU_i\rangle = | q_i \bar x\rangle^{J=1}$
\ and
\ $|\MUU_j\rangle = | q_j \bar x\rangle^{J=1}$,
\ as well as the masses of the corresponding pseudoscalars,
\ $|\MUD_i\rangle = | q_i \bar x\rangle^{J=0}$
\ and \
$|\MUD_j\rangle = | q_j \bar x\rangle^{J=0}$,
\ we have, in analogy with the second equation in (\ref{SZeq}),
\beq{mesdiff}
\langle m_{q_i} -m_{q_j} \rangle_{xMes}
=
{3(M_{\MUU_i} - M_{\MUU_j})
+(M_{\MUD_i} - M_{\MUD_j})
\over 4}=
\tilde M(V_i)- \tilde M(V_j)
\end{equation}
where $\tilde M(V_i)\equiv (3M_{\MUU_i} + M_{\MUD_i} )/4$
is the meson mass without the hyperfine contribution.

This method is not applicable to the $s \bar s$ system because of
$\eta$-$\eta^\prime$  mixing and the absence of a pseudoscalar $\bar s s$
meson.

For baryons we consider only the nucleon and the isoscalar baryons with one
heavy or strange quark, $\Lambda$, $\Lambda_c$ and $\Lambda_b$,  where the
hyperfine interaction is determined entirely by the light quark $u$ and $d$
interactions and drops out of all mass differences considered.

Thus given the masses of two baryons
\ $|B_i\rangle = | q_i ud \rangle$ \ and
\ $|B_j\rangle = | q_j ud \rangle$, \
we have
\beq{bardiff}
\langle m_{q_i} -m_{q_j} \rangle_{udBar}
= M_{q_i u d} - M_{q_j u d}
=
M_{B_i} - M_{B_j}
\end{equation}
If we now assume that
the quark mass differences from mesons (\ref{mesdiff}) are equal to
those from baryons (\ref{bardiff}) we predict that
the baryon-meson mass difference is independent of quark flavor
\beq{mesbardiff}
 M(B_i) - \tilde M(V_i) =  M(B_j) - \tilde M(V_j)\equiv \Delta MB
\end{equation}
which works to 4\%,
\begin{equation}
\begin{array}{ccccccc}
\tallstrut
 M(N) - \tilde M(\rho) &=& M(\Lambda) - \tilde M(K^*) &=&
M(\Lambda_c) -  \tilde M(D^*) &= &
M(\Lambda_b) - \tilde M(B^*)
\\
  323~\rm{MeV} &\approx& 321~\rm{MeV} &\approx&312~\rm{MeV}
 &\approx& 310~\rm{MeV}
\end{array}
\label{eq:mesbardif}
\end{equation}
The physical interpretation of this result is simple. The baryon-meson mass
differences computed above give the mass of a single constituent $u$ or $d$
quark, which is to a very good approximation independent of the flavor of
the companion quark.

We now calculate
the mass ratio \ $m_c/m_s$ \ from the hyperfine splittings
in mesons and baryons in the same way that
the mass ratio $m_s/m_u$ has been calculated\cite{SakhZel,DGG}.
We find the same value from mesons and baryons to within 1\%,
\begin{equation}
\left({{m_c}\over{m_s}}\right)_{Bar} =
{{M_{\Sigma^*} - M_\Sigma}\over{M_{\Sigma_c^*} - M_{\Sigma_c}}} = 2.84
\quad\approx\quad
\left({{m_c}\over{m_s}}\right)_{Mes} =
{{M_{K^*}-M_K}\over{M_{D^*}-M_D}}= 2.82
\end{equation}
This ratio can be applied to the doubly charmed baryons denoted by
$\Xi_{cc}$
for which there is some experimental evidence. The spin splitting
$M_{\Xi_{cc}^*} - M_{\Xi_{cc}}$ is then predicted to be
\begin{equation}
\left({{m_c}\over{m_s}}\right)_{Bar} =
{{M_{\Xi^*} - M_\Xi}\over{M_{\Xi_{cc}^*} - M_{\Xi_{cc}}}} \approx 2.8; ~ ~
~ ~ ~
M_{\Xi_{cc}^*} - M_{\Xi_{cc}} \approx 70 \hbox{\ MeV}
\end{equation}

\section{Limits of validity of the simple model}
\subsection{Problems with baryon magnetic moments}
The contrast between the successes of the simple model in a large number of
cases and its breakdown in others
can provide clues to an eventual explanation from QCD.
We now clarify
its failure
to explain the
$\Sigma$ and $\Xi$ magnetic moments\cite{hjlmag}  in
contrast with its  remarkable success in
describing the nucleon and $\Lambda$ magnetic moments and the hadron mass
spectrum \cite{ICHJLmass,DGG}.
The $SU(3)$ symmetry relations  between all
baryon octet wave functions having two quarks of the same flavor and a third
quark of a different flavor can be expressed in terms of the contributions to the
baryon spin of all quarks of flavor $u$, $d$ and $s$ denoted respectively by
$ \Delta u $, $ \Delta d $ and $ \Delta s $.

\beq{suhypa}
\Delta u (\Sigma^+) = \Delta d (\Sigma^-) = \Delta s (\Xi^o) = \Delta s (\Xi^-)
= \Delta u (p) = \Delta d (n)
\eeq
\beq{suhypb}
\Delta s (\Sigma^+) = \Delta s (\Sigma^-) =\Delta u (\Xi^o) = \Delta d (\Xi^-)
= \Delta d (p) = \Delta u (n)
\eeq

Eq.(\ref{suhypa}) states
that the contributions to  the baryon spin of the
quark pair of the same flavor are the same for all baryons; eq.(\ref{suhypb})
states that the contributions to  the baryon spin of the odd quark of different
flavor are the same for all baryons. $SU(6)$ relates these two different
contributions; $SU(3)$ does not and we do not use SU(6) here.

We assume that baryon magnetic moments  are proportional to  $\Delta u$,
$\Delta d $ and $ \Delta s $,  multiplied by the quark magnetic moments whose ratios
are given by the ratios of electric charges $(\mu_u/\mu_d) = {-}2$
and the commonly used\cite{DGG} SU(3) breaking factor $(\mu_d/\mu_s)\approx 3/2$.
Then
 baryon magnetic moments are related by the
SU(3) relations
(\ref{suhypa}) and (\ref{suhypb}) predict,

\beq{sigmanb}
{{\Delta u (\Sigma^+)}\over{\Delta u (p)}}  =
{{\mu(\Sigma^+) -\mu( \Sigma^-)}\over{2\mu (p) + \mu (n)}} = {3.6 \over 3.67} =
{{\Delta s (\Xi^o)}\over{\Delta u (p)}}\approx
{{3}\over {2}} \cdot {{-\mu(\Xi^o) - 2\mu(\Xi^-)}
\over{2\mu (p) + \mu (n)}}  ={3.82 \over 3.67} \approx 1
\eeq
\beq{xinb}
{{\Delta u (\Xi^o)}\over{\Delta d (p)}} =
{{\mu(\Xi^o) - \mu(\Xi^-)}\over{2\mu (n) + \mu(p)}} = {{0.60}\over{1.03}}
\not= 1; ~  ~ {{\Delta s (\Sigma^+)}\over{\Delta d (p)}}
\approx
{3\over 2} \cdot{{-\mu(\Sigma^+) - 2\mu(\Sigma^-)}\over{2\mu (n) + \mu(p)}} =
{{0.21}\over{1.03}}  \not= 1
\eeq
Both predictions (\ref{sigmanb})  are in excellent agreement with experiment,
while  the other two (\ref{xinb}) are in strong disagreement.

The nonstrange contributions to the $\Sigma$ moments and the
strange contributions to the $\Xi$ moments agree with the $SU(3)$ symmetry
prediction; the others do not. The contributions of
the two quarks of the same flavor in the three-quark baryon satisfy the
symmetry; those of the odd quark in hyperons are significantly less than
the odd quark contribution in the nucleon, in disagreement with the symmetry
prediction.

All our new successful relations between baryon masses involve only the $N$,
$\Lambda$ and decuplet masses, and do not involve the $\Sigma$ and $\Xi$ where
problems seem to arise.

No satisfactory understanding of this discrepancy is available at the
moment. It is an interesting challenge to use it as a clue for
identifying the correct effective degrees of freedom at low energy and
their derivation from QCD.

\subsection{Problems with hyperfine splittings}
An additional failure of the simple model arises in the
application to states with more than one heavy or
strange quark; e.g. in the experimental hyperfine
splittings in the charmed $D$ mesons\cite{pnudstar}. Experiment gives
\beq{charmd}
{{m_s}\over{m_u}} = {{M(D^*) - M(D)}\over{M(D_s^*) - M(D_s)}} =
1.01  \not=
1.60
\eeq
Some insight into the disagreement between the results
(\ref{charmd})
and (\ref{SZmag}) may be obtained by noting that
increasing a quark mass not only decreases the strength of the hyperfine
interaction but also increases the value of the wave function at the
origin and therefore increases the matrix element of the interaction.
The two effects are in opposite directions and which is dominant
is not clear a priori. The result (\ref{charmd})
suggests that the two effects may cancel in the charmed case.
But the problem remains why ignoring wave
function effects gives such good results in the lighter quark sector.
Furthermore this model has as yet no rigorous justification from QCD and
the exact meaning of constituent quarks and constituent quark masses
remain unclear.

\section*{Summary and Conclusions}

We have demonstrated a new large class of simple
phenomenological hadronic mass and
magnetic moment relations.
We obtained several new successful
relations involving hadrons containing two and three strange
quarks and hadrons containing heavy quarks
and give a new prediction regarding spin splitting between
doubly charmed baryons.
We provided numerical evidence for an effective supersymmetry between
mesons and baryons related by replacing a light antiquark by a light diquark.

The simple mass formula eq.~(\ref{sakzel}) holds with
a single set of effective quark mass values for all ground state mesons and
baryons having no more than one  strange or heavy quark and also for vector
mesons and spin 3/2 baryons having two  and three strange quarks.

The breakdown of this simple description in the $\Xi$ and $\Sigma$ magnetic
moments has been clarified. Contributions from the two quarks of the same
flavor in all baryons satisfy $SU(3)$ symmetry.
All $SU(3)$ violations are due to
suppression of the contributions from the odd strange quark in the $\Sigma$
and the odd nonstrange quark in the $\Xi$.
Direct derivation of both the successful and badly broken relations from QCD is
still an open challenge.

\section*{Acknowledgments}

The research of one of us (M.K.) was supported in part by a grant from the
Israel Science Foundation administered by the Israel Academy of Sciences and
Humanities. The research of one of us (H.J.L.) was supported in part by the
U.S. Department of Energy, Division of High Energy Physics, Contract
W-31-109-ENG-38.

%
\catcode`\@=11 
\def\references{
\ifpreprintsty \vskip 10ex
%
\hbox to\hsize{\hss \large \refname \hss }\else
\vskip 24pt \hrule width\hsize \relax \vskip 1.6cm \fi \list
{\@biblabel {\arabic {enumiv}}}
{\labelwidth \WidestRefLabelThusFar \labelsep 4pt \leftmargin \labelwidth
\advance \leftmargin \labelsep \ifdim \baselinestretch pt>1 pt
\parsep 4pt\relax \else \parsep 0pt\relax \fi \itemsep \parsep \usecounter
{enumiv}\let \p@enumiv \@empty \def \theenumiv {\arabic {enumiv}}}
\let \newblock \relax \sloppy
 \clubpenalty 4000\widowpenalty 4000 \sfcode `\.=1000\relax \ifpreprintsty
\else \small \fi}
\catcode`\@=12 
{\tighten

}
\end{document}